\documentclass[aps,prb,twocolumn,floatfix,showpacs]{revtex4}
\usepackage[dvips]{graphicx}

\begin{document}





\title{Mechanical modulation of single-electron tunneling through 
molecular-assembled metallic nanoparticles}

\author{ Yongqiang Xue $^{*}$ and Mark A. Ratner}
\affiliation{Department of Chemistry and Materials Research Center, 
Northwestern University, Evanston, Illinois 60208, USA}
\date{\today }

\begin{abstract}
We present a microscopic study of single-electron tunneling in 
nanomechanical double-barrier tunneling junctions formed using a vibrating 
scanning nanoprobe and a metallic nanoparticle connected to 
a metallic substrate through a molecular bridge. We analyze the motion of 
single electrons on and off the nanoparticle through the tunneling current, 
the displacement current and the charging-induced electrostatic force 
on the vibrating nanoprobe. We demonstrate the mechanical 
single-electron turnstile effect by applying the theory to a gold nanoparticle 
connected to the gold substrate through alkane dithiol molecular bridge 
and probed by a vibrating platinum tip.   
\end{abstract}

\pacs{85.35.Gv,85.85.+j,85.65.+h,73.63.-b}

\maketitle


\emph{Introduction.}--- Nanoelectromechanical devices that combine 
mechanics with electronics are of great interests for applications in 
electronics, precision measurement and sensors.~\cite{RC} Among 
the experimental implementations, mechanical single-electron devices 
that explore the interplay between the macroscopic motion of a 
nanomechanical element and the quantized single-electron 
transfer have attracted much attention for revealing new mechanisms 
of electron transport and schemes of mechanical detection at the 
quantum-limit.~\cite{MESET} In particular, the shuttle mechanism of 
quantized charge transfer has been proposed theoretically~\cite{Shuttle98,
ShuttleTh} and studied experimentally~\cite{ShuttleExp} 
utilizing a variety of nanomechanical elements. 
 
The model system considered in the original proposal~\cite{Shuttle98} 
for a mechanical single-electron shuttle consists of a small metal cluster 
connected to the electrodes through mechanically soft organic linkers. 
A periodic self-oscillation of the cluster in conjunction with the cluster 
charging/decharging is predicted for sufficiently large bias voltages, 
leading to an average current proportional to the self-oscillation 
frequency.~\cite{Shuttle98} Two factors neglected in the original 
proposal may complicate the analysis and prevent the observation 
of the shuttle effect in molecular-assembled single-electron 
devices:~\cite{SASET} (1) At small oscillation amplitude, the 
cluster displacement modulates the metal-molecule bond length/strength 
rather than the molecule core. The exponential dependence of 
the tunneling resistance on the displacement of the metal island 
may not hold;~\cite{XueMol1} (2) At large oscillation amplitude, 
the internal structure 
of the linker molecule may be distorted due to the forces induced by 
the change in metal-molecule bond, the net charges on the cluster and 
the applied electrical field. The shuttle mechanism of electron 
transfer has also been used to interpret recent experiment on 
a $C_{60}$ single-electron transistor,~\cite{ShuttleMol,MolTh1} but 
alternative explanations~\cite{MolTh2} also exist. Recently it has been 
suggested that measurement of shot noise spectrum~\cite{Jauho} 
and full counting statistics~\cite{Pistolesi} may help in elucidating the 
mechanism involved. 

\begin{figure}
\includegraphics[height=2.5in,width=3.0in]{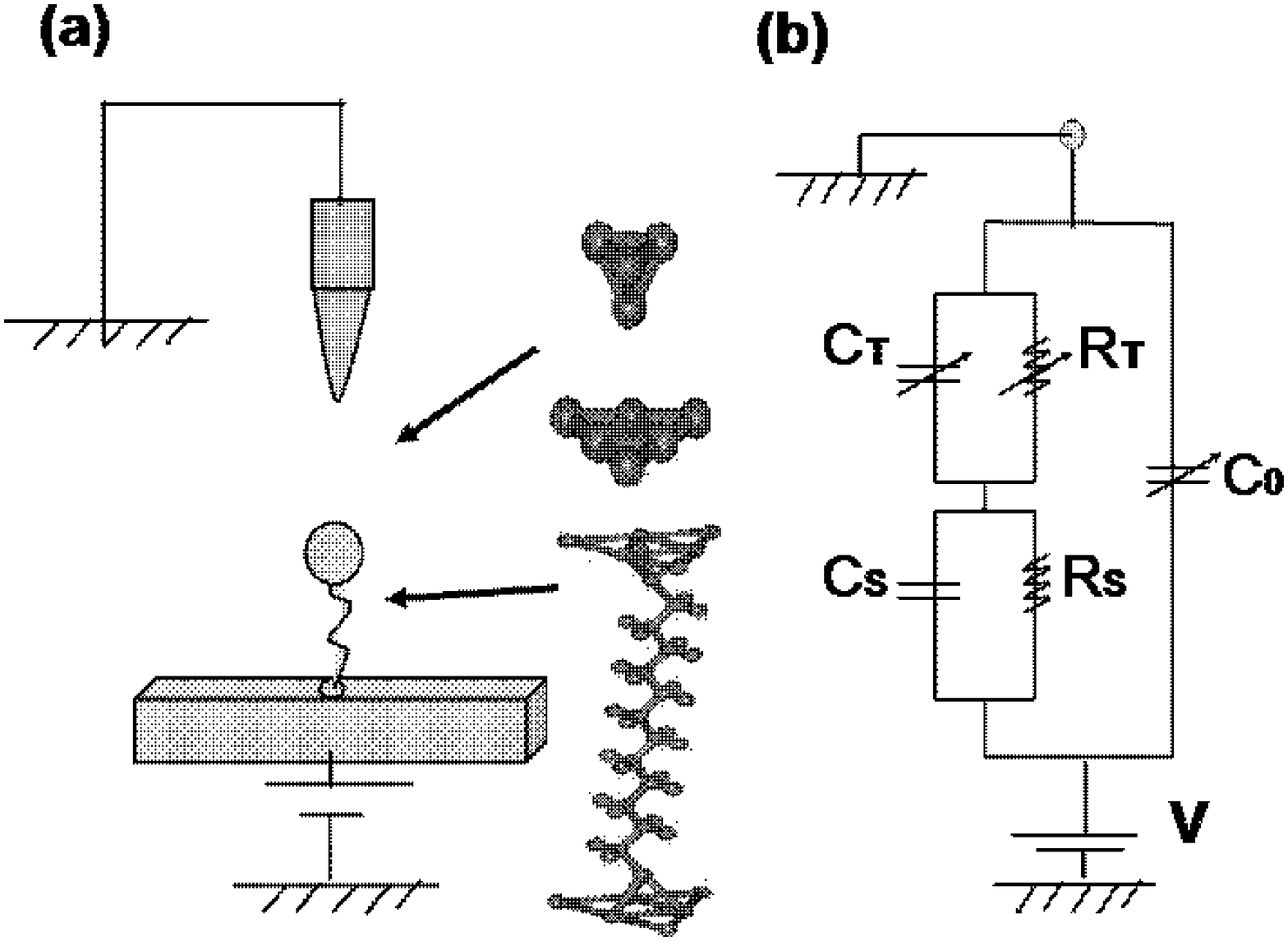}

\caption{\label{xueFig1} (Color online) (a) shows schematic 
illustration of the 
nanomechanical double-barrier tunnel junctions formed using a 
vibrating platinum tip and a gold metallic nanoparticle connected 
to the gold substrate through an alkane dithiol molecular bridge with 
12 alkane units (AK12). The platinum tip is modeled as a 4-atom 
pyramid sitting on top of the semi-infinite substrate. 
(b) shows the equivalent circuit model of the system. Note that the 
substrate junction capacitance ($C_{S}$) and resistance ($R_{S}$) 
are fixed, while the tip junction capacitance ($C_{T}$) and resistance 
($R_{T}$) are time-dependent due to the tip vibration. The tip-substrate 
capacitance ($C_{0}$) is also time-dependent.}  
\end{figure}

A simpler way of incorporating mechanical degree of freedom 
into single-electron devices is to couple one of the tunnel junctions 
to a nanomechanical oscillator,~\cite{MTunnel} using e.g., the 
microcantilever tip of a conducting Atomic Force Microscope 
(AFM) or hybrid Scanning Tunneling Microscope 
(STM)/AFM.~\cite{SPM,TipJ,TipC} 
The model system is illustrated schematically in Fig.\ \ref{xueFig1}(a), 
where the nanoparticle is connected to the substrate through an 
organic molecule (assumed rigid), while tunneling across the top 
contact can be modulated mechanically through the vibrating tip. 
The introduction of a vibrating tunnel contact leads to tunneling and 
displacement currents flowing simultaneously in accordance with the 
probe vibration,~\cite{ButtikerTD,Schoeller} which can be measured 
separately using a two-phase lock-in amplifier.~\cite{TipJ} The 
discrete-electron tunneling also induces an electrostatic force on the 
tip.~\cite{TipJ,TipC,SPM,SEF} 
Since both the tunneling/displacement currents and the force on the 
microcantilever tip can be measured as a function of the bias voltage, 
useful information regarding the interplay among tip vibration, 
discrete electron motion and metal-molecule interaction can be extracted, 
which may potentially allow applications in displacement detection and 
chemical/bio-sensing. The purpose of this paper is thus to present a 
theoretical analysis of such nanomechanical double-barrier tunneling 
junctions using realistic atomic-scale models.

\emph{Master-equation approach to single-electron tunneling through 
molecular-assembled metallic nanoparticles.}--- We consider periodic 
vibration of the tip, with the tip-nanoparticle distance being 
$x(t)=d_{0}+d_{1}cos 2\pi ft$. 
The coupled molecule-nanoparticle-vibrating tip system is described 
by the following Hamiltonian: 
\begin{equation}
\label{HALL}
H=H_{S}+H_{T}+H_{I}+V_{es}+H_{mol}+H_{T_{S}}+H_{T_{T}}, 
\end{equation}
where $H_{\alpha}=\sum_{ k } 
\epsilon_{ k \alpha}a_{ k \alpha}^{\dagger}
a_{ k \alpha} $ ($\alpha=S,T$) and  
$H_{I}=\sum_{l } \epsilon_{ l I}c_{l I}^{\dagger}c_{l I} $ describe the 
noninteracting electrons in the substrate (S), tip (T) electrodes and on 
the central island (I) respectively. We model both the electrodes and the 
nanoparticle as infinite electron reservoirs with electrochemical potential 
$\mu_{\alpha},\alpha=S,T,I$. The indices $ k$ and $l$ 
enumerate the electron states of the electrodes and the nanoparticle. 
We describe the linker molecule using an effective single-particle 
Hamiltonian $H_{mol}=\sum_{i} \epsilon_{i}b_{im}^{\dagger}b_{im}$. 
The electrostatic part of the energy is $V_{es}(q_{S},q_{T})
=\frac{q^{2}}{2C_{\Sigma}}+q_{S}V_{S}+q_{T}V_{T},
C_{\Sigma}=C_{S}+C_{T},$ where $C_{S(T)}$ and $q_{S(T)}$ are the 
capacitance and charge of the substrate (tip) junction 
respectively.~\cite{AKBook} The voltage drop across the 
substrate (tip) junction is determined from the tip-substrate 
bias voltage $V$ by $V_{S(T)}=\frac{C_{T(S)}}{C_{\Sigma}}V$. 
The net charge on the central island is $q=q_{S}-q_{T}=-e(n+n_{x})$, 
where $en_{x}$ is the background charge and $en$ is the quantized 
island charge.  Unlike the 
conventional single-electron devices where the background charge is 
typically induced by charged impurities embedded in the insulating layer, 
here an intrinsic background charge $n_{x}$ can be induced by the 
charge transfer between the molecule and the central island during the 
formation of the substrate junction,~\cite{XueNP03} and can be obtained 
from microscopic molecular junction calculations.~\cite{XueMol2} 

\begin{figure}
\includegraphics[height=2.5in,width=3.0in]{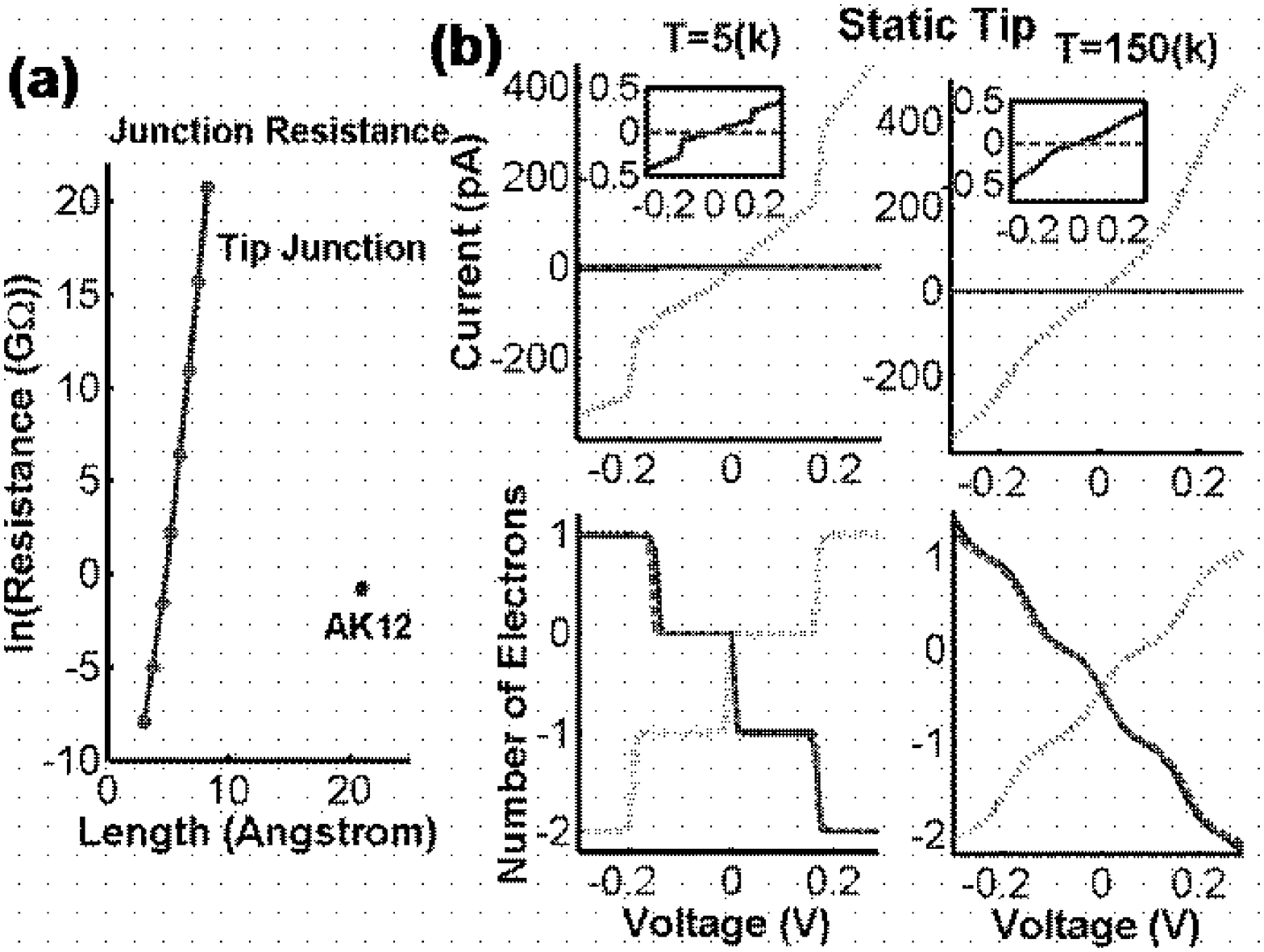}
\caption{\label{xueFig2} (a) shows the molecular junction and 
tip junction resistance as a function of the junction length. The tip 
junction resistance increases exponentially with tip-nanoparticle 
distance as the tip remains weakly coupled to the nanoparticle during 
the tip vibration cycle.  
(b) shows the static current-voltage (I-V) and excess island 
charge-voltage (n-V) characteristics of the double-barrier tunnel junctions 
for a 10(nm)-diameter nanoparticle at tip-nanoparticle distance of 
$3.0 \AA $ (dotted line), $6.0 \AA $ (solid line) and $9.0 \AA $ (dashed 
line) for temperatures of $T=5,150 K$ respectively. The inset shows the 
magnified view of I-V characteristics at $6.0 \AA $ and $9.0 \AA $. Note 
that at $x=6.0 \AA $, we already have $R_{T} >> R_{S}$, so the 
n-V characteristics remains approximately the same as $x$ increases 
further. }
\end{figure}

\begin{figure}
\includegraphics[height=2.5in,width=3.0in]{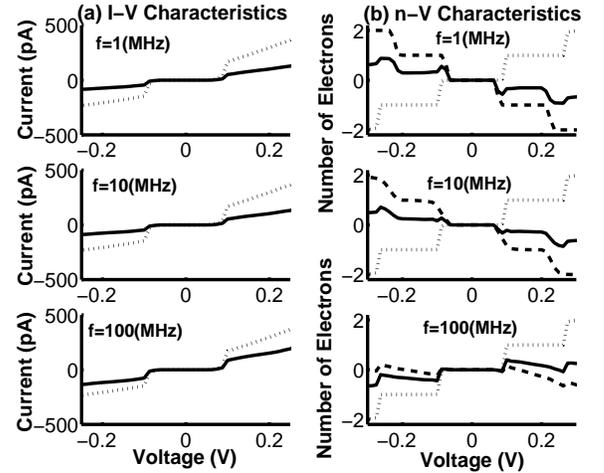}
\caption{\label{xueFig3} (a) shows  
the time-averaged tunneling I-V characteristics (solid line) and the 
instantaneous tunneling I-V characteristics at $x=3.0 \AA $ (dotted line) 
at three tip vibrating frequencies of $f=1,10,100 MHz$.  
The tunnel currents are virtually independent of the tip vibrating frequency.  
(b) shows the time-averaged n-V characteristics (solid line), the 
instantaneous n-V characteristics at $x=3.0 \AA $ (dotted line) and 
$x=9.0 \AA $ (dashed line) respectively. The instantaneous values 
are the values obtained at the moments when the tip moves to 
$x=3.0 \AA $ or $9.0\AA $  }
\end{figure}

\begin{figure}
\includegraphics[height=2.5in,width=3.0in]{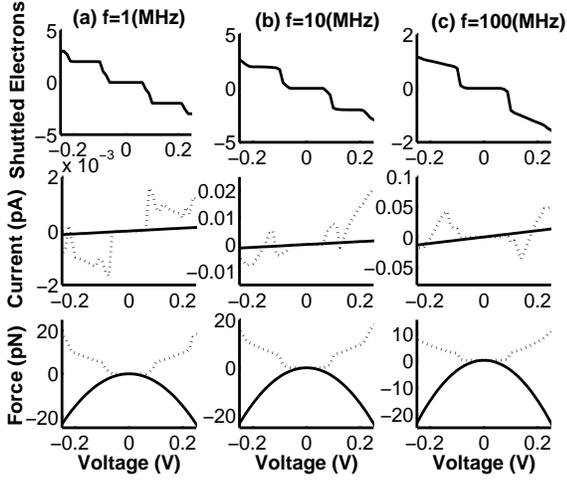}
\caption{\label{xueFig4} Upper figures show the average number of 
shuttled electron per cycle of tip vibration, middle figures show the 
average displacement current, and lower figures show the average 
electrostatic force on the metallic tip as a function of tip-substrate 
bias voltage at tip vibrating frequencies of $f=1,10,100(MHz)$. 
In the middle figures, we have shown both the capacitive part 
(solid line) and ``Coulomb Blockade'' part (dotted line) 
of the displacement current. The capacitive part increases linearly 
with voltage and tip vibrating frequency. But the ``Coulomb Blockade'' 
part oscillates with increasing voltage. In the lower figures, we have 
shown both the capacitive part $F_{Cap}$ (solid line) and discrete-electron 
part  $F_{CB}$ (dotted line) of the electrostatic force on the tip. Note 
that the step-wise increase in the shuttled electrons in the upper 
figures correlates with the step-like increase in discrete-electron force 
$F_{CB}$ in the lower figures. }

\end{figure}

The transfer of single-electrons is mediated by tunneling through the 
molecular (substrate) junction 
$H_{T_{S}}  =  \sum_{ik}  (t_{im;kS}b_{im}^{\dagger}a_{kS}
 +t_{im;lI}b_{im}^{\dagger}c_{lI}e^{-i \phi_{S}} + H.C.)$,  
and the tip junction
$H_{T_{T}}(t)  =  \sum_{lk} 
 (t_{lI;kT}(x(t))a_{kT}^{\dagger}c_{lI}e^{-i \phi_{T}} + H.C.)$, 
where we have shown explicitly the time-dependence of the tip-nanoparticle 
coupling through the tip position $x(t)$. The phase 
operator $\phi_{S(T)}$, canonically conjugate to the charge $q_{S(T)}$ on 
the substrate (tip) junction, keeps track of the quantized electron tunneling 
since $[\phi_{S(T)},q_{S(T}]=ie$ and 
$e^{i\phi_{S(T)}}q_{S(T)}e^{-i\phi_{S(T)}}
=q_{S(T)}-e$.~\cite{INBook,Schon} The electrostatic energy change 
after tunneling of an electron through the substrate(tip) junction is thus 
$\Delta_{n}=e^{i\phi_{S(T)}}V_{es}e^{-i\phi_{S(T)}}-V_{es}=eV_{S(T)}
+\frac{(q-e)^{2}}{2C_{\Sigma}}
-\frac{q^{2}}{2C_{\Sigma}}$.~\cite{AKBook,INBook} 

We describe single-electron transport using the master 
equation:~\cite{ShuttleTh,Schoeller,Niu97}
\begin{eqnarray}
\label{Rate}
\frac{dP_{m}(t)}{dt} &= & P_{m+1}(t)\Gamma^{-}(\Delta_{m},t)
                             +P_{m-1}(t)\Gamma^{+}(\Delta_{m-1},t) \nonumber \\
         &-&P_{m}(t)(\Gamma^{-}(\Delta_{m-1},t)+\Gamma^{+}(\Delta_{m},t))
\end{eqnarray}
where $P_{m}$ is the probability of finding $m$ excess electrons on the 
nanoparticle and 
$\Gamma^{+(-)}=\Gamma^{+(-)}_{S}+\Gamma^{+(-)}_{T}$. 
The rate for transition $m \to m\pm 1$ due to electron tunneling through the 
substrate (tip) junction, $\Gamma_{S(T)}^{+(-)}$, is obtained from the 
electron transmission coefficient of the corresponding junctions 
using
$\Gamma_{S(T)}^{+}(w,t) =  \frac{2}{ h} 
  \int dE T_{S(T)}(E,V;t)f_{S(T)}(E)(1-f_{I}(E-w)) $ 
and $  \Gamma_{S(T)}^{-}(w,t) = \frac{2}{h} 
  \int dE T_{S(T)}(E,V;t)(1-f_{S(T)}(E))f_{I}(E-w) $,~\cite{XueNP03}  
where $w=\Delta_{m}$ or $\Delta_{m-1}$. The transmission 
coefficient $T_{S(T)}$ can be  
determined from the standard Non-Equilibrium Green's Function 
theory~\cite{NEGF} as developed for molecular tunnel 
junctions.~\cite{XueNP03,XueMol2} Here the time dependence 
in the tunneling rate $\Gamma^{+(-)}$ is introduced through the 
time-dependent tunneling rate across the tip junction 
$\Gamma^{+(-)}_{T}$ due to the tip vibration $x(t)$. Since the 
tunneling time for electron tunneling across the tip junction is much 
smaller than the tip vibration period, the time-dependence in $T_{T}$ 
indicates that we calculate the transmission coefficient using the tunneling 
Hamiltonian $H_{T_{T}}(t)$ at the instantaneous tip position $x(t)$. 
   
The tunneling current across the tip junction and the average excess 
electron number on the nanoparticle at time $t$ are given by 
$I_{T}^{Tun}(t) = e \sum_{m} P_{m}(t)
[\Gamma^{+}_{T}(\Delta_{m},t)-\Gamma^{-}_{T}(\Delta_{m-1},t)]$ 
and $n(t) = \sum_{m}mP_{m}(t)$ respectively. 
The displacement current flowing through the vibrating tip is due to 
time-variation of system charges and is fixed by electrostatic 
considerations~\cite{Schoeller,INBook}: 
\begin{eqnarray}
\label{IDis}
I_{T}^{Dis}(t) &=& \dot{C}_{0}V+\frac{dq_{T}}{dt} 
              =I_{Cap}(t)+I_{CB}(t) \nonumber \\
             &=& \frac{d}{dt} (C_{0}+\frac{C_{S}C_{T}}{C_{S}+C_{T}})V 
          +e\frac{d}{dt}(\frac{C_{T}}{C_{S}+C_{T}}(n(t)+n_{x}))
\end{eqnarray}
where $C_{0}$ is the capacitance between the tip and the substrate. 
The first term $I_{Cap}$ is the displacement current induced by the direct 
capacitive coupling between the tip and the substrate and the series 
capacitance of the double-barrier junction. The second term 
$I_{CB}$ gives the ``Coulomb blockade''  component of displacement current  
due to the time variation of the excess electrons on the nanoparticle. 
The electrostatic force on the oscillating tip can also be 
divided into the capacitive force $F_{Cap}$ and the discrete-electron 
force $F_{CB}$ as follows:
\begin{eqnarray}
\label{FES}
F_{tip} &=& -\frac{1}{2}\frac{dC_{0}}{dx}V^{2}  
    -\frac{1}{2}\frac{d(q_{T}V_{T})}{dx}  = F_{Cap}+F_{CB} \nonumber \\ 
&=&-\frac{1}{2}\frac{d}{dx}(C_{0}
  +C_{T}(\frac{C_{S}}{C_{\Sigma}})^{2})V^{2} 
      -\frac{e}{2}\frac{d}{dx}
(\frac{C_{S}C_{T}}{C_{\Sigma}^{2}}(n+n_{x}))V. 
\end{eqnarray}
Since our main interests here are in electron transport and 
charging-induced forces, we neglect 
non-voltage dependent forces like the long-range van der Waals force. 
Since we consider transport within the 
Coulomb blockade regime, the tip remains weakly coupled to the 
nanoparticle during the entire cycle of vibration, and we neglect also 
the short-range contact force that may arise from the 
tip-nanoparticle bonding. Both type of forces would have been 
important if we want to study the detailed oscillation dynamics of the 
cantilever.~\cite{SPM}  

\emph{Device model.}--- We apply the theory to single-electron tunneling 
through a 10(nm)-diameter gold nanoparticle connected to the gold 
$\langle 111 \rangle$ substrate through an alkane dithiol 
molecule with 12 alkane units (AK12) and probed using a vibrating 
platinum tip. Electron transport through gold-dithiol 
molecule-gold junctions have been recently studied in detail using a 
first-principles self-consistent matrix Green's function 
theory,~\cite{XueMol2} from which we obtain the intrinsic background 
charge $en_{x}$ induced by the charge transfer from the molecule to the 
nanoparticle and the transmission coefficient for electron tunneling through 
the metal-molecule-nanoparticle junction. To summarize, 
the gold electrodes are modeled as semi-infinite $\langle 111 \rangle$ 
single-crystals. Six nearest-neighbor gold atoms on each metal 
surface (twelve gold atoms overall) are included into the 
``extended molecule'' where the self-consistent calculation is performed. 
The rest of the electrodes (with the six atoms on each surface removed) 
are considered as infinite electron reservoirs, whose effects are included 
as self-energy operators. The calculation is performed 
using the Becke-Perdew-Wang parameterization of density-functional 
theory ~\cite{BPW91} and appropriate pseudopotentials with 
corresponding optimized Gaussian basis sets,~\cite{XueMol2} We find 
that the resistance of the AK12 molecular junction is 
$R_{S} \approx 300(M\Omega)$, and the background charge 
is $n_{x}\approx -0.45$.~\cite{Note}  

Since the tip geometry in STM/AFM is often not well characterized, we 
model the Pt tip as a 4-atom Pt pyramid sitting on top of the 
$\langle 111 \rangle$ Pt substrate. The tunneling matrix 
elements are obtained using the semi-empirical Extended Huckel 
Theory (EHT)~\cite{HoffXue} and considering coupling between 
the Pt pyramid and six neighbor gold atoms on the surface of the gold 
nanoparticle, with the apex Pt atom sitting in front of the center of the six 
gold atoms arranged following the $\langle 111 \rangle$ gold surface 
geometry (Fig. \ref{xueFig1}(a)). The calculated tip junction resistance 
$R_{T}$ as a function of the tip apex-nanoparticle distance is shown in 
Fig. \ref{xueFig2}(a), which increases exponentially with the 
tip-nanoparticle distance once it falls within the weak-coupling regime. 
For capacitance modeling, such atomic-scale model is not needed. The tip 
is instead replaced by a $1 \mu m$-radius sphere, which takes into 
account approximately the tip curvature effect and is common in 
AFM research.~\cite{SPM,TipCap} Here both the 
substrate and tip junction capacitance are obtained from classical 
electrostatics of conducting sphere sitting in front of the conducting 
substrate,~\cite{TipCap} which gives the substrate junction capacitance 
of $0.9 aF$, the tip junction capacitance of $0.97 aF$ and 
tip-substrate capatitance of $0.34 fF$ at tip-nanoparticle distance 
$x=9.0 \AA$ respectively.

\emph{Results and their interpretation.}--- We assume the tip vibration  
being described by oscillation amplitude of $d_{1}=3.0 \AA $ and average 
tip-nanoparticle distance of $d_{0}=6.0 \AA $. The device functions as a 
nanomechanical single-electron turnstile, which can be understood by 
examining the static (tip not moving) excess island electron-voltage (n-V) 
characteristics at different tip-nanoparticle distances 
$x$ (Fig. \ref{xueFig2} (b)). We find that the trend of the 
charging state on the 
nanoparticle is reversed as the tip-nanoparticle $x$ increases from 
$3.0 \AA$ to $9.0 \AA$. This is because at $x=3.0 \AA$, the tip junction 
resistance is much smaller than the substrate junction resistance 
while at $x=9.0 \AA$, the tip junction resistance becomes much larger 
than the substrate junction resistance (Fig. \ref{xueFig2} (a)). 
As the tip starts vibrating at $x=d_{0}+d_{1}=9.0 \AA $, 
the nanoparticle will be charged by discrete-electron tunneling 
across the substrate junction if the bias voltage is large enough 
to overcome the charging energy. As the tip moves close to the 
nanoparticle at $x=d_{0}-d_{1}=3.0 \AA $, the tip junction 
becomes sufficiently conductive ($R_{T} << R_{S}$) allowing 
the stored excess charge on the nanoparticle to be transferred 
onto the tip. A discrete number of electrons 
is thus being shuttled across the system per tip vibrating cycle. 
Note that the n-V characteristics at $x=d_{0}=6.0 \AA$ is similar to 
that at $x=9.0 \AA$ because $R_{T}$ is already much larger than 
$R_{S}$ at $x=6.0 \AA$, which is also the case during most part of the 
tip vibration cycle. 

We solve the rate equation (Eq.\ \ref{Rate}) numerically,~\cite{KKL} from 
which we obtain the time-dependent tunneling current, the displacement 
current and the electrostatic force on the Pt tip. We calculate the probability 
$P_{\Delta}(m)$ that m electrons have been transferred during one 
cycle of tip vibration from 
$P_{\Delta}(m)=\sum_{k}P_{i}(k)P(k-m|k,t_{0}+1/f)$ (f is the vibrating 
frequency),~\cite{ShuttleTh} 
where $P_{i}(k)$ is the possibility of finding k electrons on the nanoparticle 
at time $t_{0}$. $P(k-m|k,t_{0}+1/f)$ is the conditional probability of 
finding $k-m$ electrons on the nanoparticle after one cycle of tip 
oscillation if there are $k$ electrons at initial time $t_{0}$, which is 
obtained by solving the rate equation Eq.\ \ref{Rate} with the initial 
condition $P(k,t_{0})=1,P(i,t_{0})=0,i \ne k$.~\cite{ShuttleTh}  
The average number of electrons being shuttled across the system is  
$n_{shuttle}=\sum_{m}mP_{\Delta}(m)$. 
       
The average  (time-average over one cycle of the tip vibration) tunneling 
current-voltage (I-V) flowing across the tip junction and excess island 
electron-voltage (n-V) characteristics of the vibrating double-barrier 
junctions at temperature of $5 K$ are shown in 
Fig. \ref{xueFig3} for three tip vibrating frequencies $f=1,10,100(MHz)$. 
For comparison, we have also shown the instantaneous tunneling 
I-V characteristics  when the tip is closest to the nanoparticle 
($x=3.0 \AA $) in Fig.\ \ref{xueFig3}(a), the instantaneous 
n-V characteristics at two different tip-nanoparticle distances of 
$x=3.0$, and $9.0 \AA $ in Fig.\ \ref{xueFig3}(b). 
The instantaneous values at $x=3.0 \AA, 9.0 \AA$ are the 
values obtained at the moment when the tip moves to 
$x=3.0\AA, 9.0 \AA $ respectively. Since the 
instantaneous tunneling current of the vibrating double-barrier junction 
decreases rapidly as tip moves away from the nanoparticle, and the tip 
velocity approaches zero as tip moves 
closer to the nanoparticle, the average tunneling I-V characteristics is 
similar in both magnitude and voltage dependence to the instantaneous 
tunneling I-V characteristics at $x=3.0\AA $ at all three tip 
frequencies. Note that the tunneling currents shown here are virtually 
independent of the tip vibrating frequency, since the time it takes  
for electron to tunnel across the tip junction is the smallest of all 
time scales involved here and can be considered as instantaneous 
compared to the tip vibration. 

The n-V characteristics show stronger tip-frequency dependence than 
the tunneling I-V characteristics. The instantaneous n-V characteristics 
at $x=3.0 \AA $ changes little with tip frequency. They are also 
similar to the static n-V characteristics in Fig. \ref{xueFig2}(b). 
This is because at x=3 \AA, tip junction resistance is small. The 
corresponding RC charging time is much smaller than the tip vibrating 
period, the instantaneous charging state on the nanoparticle is thus not 
affected by the tip vibration. As the tip moves away from the 
nanoparticle, $R_{T}$ increases rapidly leading to larger $RC$ charging 
time. Therefore charge imbalance on the nanoparticle can be affected 
at large tip-nanoparticle distance as the tip vibrating frequency increases, 
which is seen clearly from the instantaneous n-V characteristics at the 
largest tip-nanoparticle distance $x=9.0 \AA $. The time-averaged 
n-V characteristics approximates well the average between the 
instantaneous n-V characteristics at $x=3.0 \AA $ and $9.0 \AA $. 

Fig. \ref{xueFig4} shows the number of electrons $n_{shuttle}$ 
being shuttled across the system per cycle of tip vibration, the 
displacement current and the electrostatic force on the tip averaged over 
one cycle of tip vibration at tip vibrating frequencies of $f=1,10,100 MHz$ 
respectively. Note that $n_{shuttle}$ decreases with 
increasing tip frequency due to reduced efficiency of charge transfer from 
the central island to the tip. The capacitive part of the displacement 
current $I_{Cap}$ increases linearly with 
the tip-substrate bias voltage (solid line), but the ``Coulomb Blockade'' 
part of the displacement current $I_{CB}$  
oscillates with the bias voltage, since $dn/dt$ changes sign during 
one cycle of the tip vibration. The peak and dip positions in $I_{CB}$ 
correlate closely with the step-wise change in the shuttled electrons with 
respect to the bias voltage. The magnitude of $I_{Cap}$ increases linearly 
with increasing tip vibration frequency, but the increase in magnitude of  
$I_{CB}$ is less regular. We have also shown the time-averaged 
capacitive force $F_{Cap}$ and the discrete-electron force $F_{CB}$ 
on the tip separately in Fig. \ref{xueFig4}. Note that $F_{Cap}$ is 
quadratic voltage-dependent and is attractive. In contrast, the 
discrete-electron force $F_{CB}$ is repulsive. The magnitude of 
$F_{CB}$ shows step-wise increase with bias voltage as more electrons 
are allowed to be shuttled, similar to the voltage-dependence of 
$n_{shuttle}$.~\cite{TipC} The overall magnitude of $F_{Cap}$ 
and $F_{CB}$ is comparable. Unlike the displacement current, 
the magnitude 
of the discrete-electron force depends only weakly on the tip frequency 
through the frequency dependence of $n_{shuttle}$. 
 
To conclude, we have presented a microscopic theory of mechanical 
modulation of single-electron tunneling through a molecular-assembled 
metal nanoparticle induced by a vibrating nanoprobe. The effects 
discussed here may be of interest for applications in 
nanomechanical chemical/bio-sensors. 
    
This work was supported by the DARPA Moletronics program, 
the DoD-DURINT program and the NSF Nanotechnology Initiative.


\begin{references}
\bibitem[*]{Xue} Author to whom correspondence should be addressed. 
Email: ayxue@chem.northwestern.edu 
\bibitem{RC} M.L. Roukes, Phys. World {\bf 14}(2), 25 (2001); 
A. Cho, Science {\bf 299}, 36 (2002). 
\bibitem{MESET} A.N. Cleland and M.L. Roukes, Nature {\bf 392}, 160 
(1998); R.E.S. Polkinghorne and G.J. Milburn, Phys. Rev. A {\bf 64}, 
42318 (2001); A.D. Armour and M.P. Blencowe, 
Phys. Rev. B {\bf 64}, 353111 (2001); 
A.D. Armour, M.P. Blencowe, and K.C. Schwab, 
Phys. Rev. Lett. {\bf 88}, 148301 (2002); A.N. Cleland, Nature {\bf 424}, 
291 (2003); M.D. LaHaye, O. Buu, B. Camatota, and K.C. Schwab, Science 
{\bf 304}, 74 (2004); A. D. Armour, M. P. Blencowe, and Y. Zhang, 
Phys. Rev. B {\bf 69}, 125313 (2004). 
\bibitem{Shuttle98} L.Y. Gorelik, A. Isacsson, M.V. Voinova, B. Kasemo, 
R.I. Shekhter, and M. Jonson, Phys. Rev. Lett. {\bf 80}, 4526 (1998).  
\bibitem{ShuttleTh} C. Weiss and W. Zwerger, Europhys. Lett. {\bf 47}, 
97 (1999); N. Nishiguchi, Phys. Rev. B {\bf 65}, 35403 (2001); T. Nord, 
L.Y. Gorelik, R.I. Shekhter, and M. Jonson, \emph{ibid.} 
{\bf 65}, 165312 (2002); A.D. Armour and A. Mackinnon, 
\emph{ibid.} {\bf 66}, 35333 (2002); 
T. Novotn{\'y}, A. Donarini, and A.-P. Jauho, 
Phys. Rev. Lett. {\bf 90}, 256801 (2003); D. Fedorets, L. Y. Gorelik, 
R. I. Shekhter, and M. Jonson, \emph{ibid.} {\bf 92}, 166801 (2004). 
\bibitem{ShuttleExp} M.T. Tuominen, R.V. Krotkov, and M.L. Breuer, 
Phys. Rev. Lett. {\bf 83}, 3025 (1999); A. Erbe, C. Weiss, W. Zwerger, 
and R.H. Blick, Phys. Rev. Lett. {\bf 87}, 96106 (2001).
\bibitem{SASET} J.R. Petta, D.G. Salinas, and 
D.C. Ralph, Appl. Phys. Lett. {\bf 77}, 4419 (2000).  
\bibitem{XueMol1} Y. Xue and M.A. Ratner, Phys.\ Rev.\ B {\bf 68}, 
115407 (2003).   
\bibitem{ShuttleMol} H. Park, J. Park, A.K.L. Lim, E.H. Anderson, 
A.P. Alivisatos, and P.L. McEuen, Nature {\bf 407}, 57 (2000).  
\bibitem{MolTh1} D. Dedorets, L.Y. Gorelik, R.I. Shehter, and 
M. Jonson, Europhys. Lett. {\bf 58}, 99 (2002). 
\bibitem{MolTh2} D. Boese and H. Schoeller, Europhys. Lett. {\bf 54}, 
668 (2001); K.D. McCarthy, N. Prokof'ev, and M. T. Tuominen, 
Phys. Rev. B {\bf 67}, 245415 (2003).
\bibitem{Jauho} T. Novotn{\'y}, A. Donarini, C. Flindt, 
and A.-P. Jauho, Phys. Rev. Lett. {\bf 92}, 248302 (2004). 
\bibitem{Pistolesi} F. Pistolesi, Phys. Rev. B {\bf 69}, 245409 (2004). 
\bibitem{MTunnel} N.F. Schwabe, A.N. Cleland, M.C. Cross, 
and M.L. Roukes, Phys. Rev. B {\bf 52}, 12911 (1995). 
\bibitem{TipJ} K. Nagano, A. Okuda, and Y. Majima, Appl. Phys. Lett. 
{\bf 81}, 544 (2002); Y. Majima, K. Nagano, and A. Okuda, 
Jpn. J. Appl. Phys. {\bf 41}, 5381 (2002).
\bibitem{TipC} Y. Suganuma, P.-E. Trudeau, and A.-A. Dhirani, 
Phys. Rev. B {\bf 66}, 241405 (2002); Y. Suganuma, 
P.-E. Trudeau, A.-A. Dhirani, B. Leathem, and B. Shieh, 
J. Chem. Phys. {\bf 118}, 9769 (2003). 
\bibitem{SPM} F.J. Giessibl, Rev. Mod. Phys. {\bf 75}, 949 (2003).
\bibitem{ButtikerTD} M. Buttiker, J. Low. Temp. Phys. {\bf 118}, 
519 (2000); B. Wang, J. Wang, and H. Guo, Phys. Rev. Lett. 
{\bf 82}, 398 (1999). 
\bibitem{Schoeller} C. Bruder and H. Schoeller, Phys. Rev. Lett. {\bf 72}, 
1076 (1994). 
\bibitem{SEF} C. Sch{\"o}nenberger and S.F. Alvarado, Phys. Rev. 
Lett. {\bf 65}, 3162 (1990); L.J. Klein and C.C. Williams, 
Appl. Phys. Lett. {\bf 79}, 1828 (2001). 
\bibitem{AKBook} D.V. Averin and K.K. Likharev, in 
\emph{Mesocopic Phenomena in Solids}, edited by B.L. Altshuler, 
P.A. Lee and R.A. Webb (Elsevier, Amsterdam, 1991). 
\bibitem{XueNP03} Y. Xue and M.A. Ratner, Phys. Rev. B {\bf 68}, 235410 
(2003); Mater. Res. Soc. Symp. Proc. {\bf 735}, C5.5 (2003).
\bibitem{XueMol2} Y. Xue, S. Datta and M.A. Ratner, Chem. Phys. 
{\bf 281}, 151 (2002); Y. Xue and M.A. Ratner, Phys.\ Rev.\ B {\bf 68}, 
115406 (2003); {\bf 69}, 85403 (2004). 
\bibitem{INBook} G.-L. Ingold and Yu.V. Nazarov, in \emph{Single Charge 
Tunneling}, edited by H. Grabert and M.H. Devoret (Plenum, New York, 
1992). 
\bibitem{Schon} H. Schoeller and G. Sch\"{o}n, Phys. Rev. B {\bf 50}, 
18436 (1994); J. K\"{o}nig, H. Schoeller and G. Sch\"{o}n, 
\emph{ibid.} {\bf 58}, 7882 (1998).   
\bibitem{Niu97} C. Liu and Q. Niu, Phys. Rep. {\bf 286}, 349 (1997). 
\bibitem{NEGF} Y. Meir and N.S. Wingreen, Phys. Rev. Lett.\ {\bf 68}, 
2512 (1992);  A.P. Jauho, N.S. Wingreen and Y. Meir, Phys. Rev. B 
{\bf 50}, 5528 (1994); H. Haug and A-P. Jauho, \emph{Quantum Kinetics in 
Transport and Optics of Semiconductors} (Springer-Verlag, Berlin, 1996).  
\bibitem{BPW91} A.D. Becke, Phys.\ Rev.\ A.\ {\bf 38}, 3098 (1988); 
M. Ernzerhof, J.P. Perdew and K. Burke, in \emph{Density 
Functional Theory I}, edited by R. F. Nalewajski (Springer, Berlin, 1996). 
\bibitem{Note} Since the metal-molecule interaction is a local 
phenomenon, we neglect the curvature of the nanoparticle surface 
and calculate the electronic structure of metal-molecule-nanoparticle 
junction in the same way as that of the metal-molecule-metal junction. 
The transferred charge and therefore the background charge $n_{x}$ are  
obtained by integrating the (self-consistent) electron density distribution 
over the region occupied by the perturbed surface 
atoms of the nanoparticle.  
\bibitem{HoffXue} R. Hoffmann, Rev.\ Mod.\ Phys.\ {\bf 60}, 601 (1988); 
Y. Xue, S. Datta, S. Hong, R. Reifenberger, J.I. Henderson 
and C.P. Kubiak, Phys. Rev. B {\bf 59}, 7852 (1999).
\bibitem{TipCap} Y. Oyama, Y. Majima, and M. Iwamoto, J. Appl. Phys. 
{\bf 86}, 7087 (1999); C. Wasshuber, \emph{Computational 
Single-Electronics} (Springer-Verlag, Wien, 2001). 
\bibitem{KKL} L.R.C. Fonseca, A.N. Korotkov, K.K. Likharev, and 
A.A. Odintsov, J. Appl. Phys. {\bf 78}, 3238 (1995). 
\end{references}
\end{document}